\documentclass[11pt,a4paper,twoside,groupcitations]{article}
\usepackage[T1]{fontenc}
\usepackage[ansinew]{inputenc}
\usepackage[english]{babel}
\usepackage{amsfonts}
\usepackage{amsmath}
\usepackage{array}
\usepackage{amsthm}
\usepackage{amssymb}
\usepackage{graphicx}
\usepackage{subfigure}
\usepackage{braket}
\usepackage{eucal}
\usepackage{verbatim}
\usepackage[table]{xcolor}
\usepackage{caption}
\usepackage{cite}
\usepackage{textcomp}
\usepackage{cancel,soul,ulem}
\usepackage{multicol}
\usepackage{tikz}
\usetikzlibrary{positioning,arrows}
\usetikzlibrary{decorations.pathmorphing}
\usetikzlibrary{decorations.markings}
\usetikzlibrary{calc,decorations.markings}
\usetikzlibrary{arrows,shapes}
\usetikzlibrary{matrix,arrows}
\usepackage{pgfplots}
\usepackage{xparse}
\definecolor{jade}{HTML}{00A86B}
\newcommand{\be}{\begin{eqnarray}}
\newcommand{\ee}{\end{eqnarray}}

\renewcommand{\d}{\mbox{${\rm d}$}} 
\newcommand{\lp}{\ell_{\rm p}}
\newcommand{\mpl}{m_{\rm p}}
\newcommand{\gn}{G_{\rm N}}
\newcommand{\rh}{r_{\rm H}}
\newcommand{\Rh}{R_{\rm H}}

\title{\bf Compact sources and cosmological horizons in lower dimensional bootstrapped Newtonian gravity}
\author{Roberto~Casadio$^{ab}$\thanks{E-mail: casadio@bo.infn.it},
$\ $
Octavian~Micu$^c$\thanks{E-mail: octavian.micu@spacescience.ro}
$\ $
and
Jonas~Mureika$^{d}$\thanks{E-mail: jmureika@lmu.edu}
\\
\\
$^a${\it Dipartimento di Fisica e Astronomia, Universit\`a di Bologna}
\\
{\it via Irnerio~46, 40126 Bologna, Italy}
\\
\\
$^b${\it I.N.F.N., Sezione di Bologna, I.S.~FLAG}
\\
{\it viale B.~Pichat~6/2, 40127 Bologna, Italy}
\\
\\
{\it $^b$Institute of Space Science, Bucharest, Romania}
\\
{\it P.O. Box MG-23, RO-077125 Bucharest-Magurele, Romania}
\\
\\
{\it $^d$Department of Physics, Loyola Marymount University}
\\
{\it Los Angeles, California, USA}
}
\begin{document}
\maketitle
\begin{abstract}
\noindent We study the bootstrapped Newtonian potential generated by a localised source in one and two
spatial dimensions, and show that both cases naturally lead to finite spatial extensions of the
outer vacuum.
We speculate that this implies the necessary existence of a cosmological (particle) horizon
associated with compact sources.
In view of the possible dimensional reduction occurring in ultra-high energy processes --
like scatterings at Planckian energies, the gravitational collapse of compact objects or the
end-point of black hole evaporation --  
one can consider such lower-dimensional ``bubbles'' immersed in our Universe as 
describing (typically Planckian size) baby universes relevant to those dynamics.
\par
\null
\par
\noindent
\textit{PACS - 04.70.Dy, 04.70.-s, 04.60.-m}
\end{abstract}
\section{Introduction and motivation}
\setcounter{equation}{0}
\label{S:intro}
The prospect of studying gravitation in lower dimensions, namely $(1+1)$
and $(1+2)$-dimensional spacetimes, began as a pedagogical 
curiosity~\cite{gott1,gott2,jackiw,rbm11,robb2,Mann:1991qp,btz,grumiller,mann2,mann3}.
It was later noted that such solutions presented  new avenues to quantizing gravity,
particularly in the $(1+1)$-dimensional case.
More recently, high energy dimensional reduction has been found to be a natural property
of disparate approaches to quantum gravity, including loop quantum gravity~\cite{lqg}, 
causal dynamical triangulations~\cite{cdts}, asymptotically safe
gravity~\cite{LaR05,ReS11}, noncommutative geometry~\cite{MoN10},
multi-fractal geometry~\cite{Cal12}, and modified dispersion relations~\cite{AAG13+}.
\par
Mainly motivated by the need to remove the singularities predicted by General Relativity~\cite{HE,geroch}
and inspired by corpuscular models of black holes and cosmology~\cite{Giusti:2019wdx,DvaliGomez,dvaliC},
bootstrapped Newtonian gravity is another attempt at describing quantum features of gravity in the strong
field regime where the graviton self-interactions cannot be
discarded~\cite{Casadio:2017cdv,Casadio:2016zpl,BootN,Casadio:2019cux,Casadio:2019pli,
Casadio:2019tfz,Casadio:2020ueb,Casadio:2020mch,Casadio:2020kbc}.
In particular, this approach starts from Deser's conjecture~\cite{deser} that it should be possible
to obtain the dynamics of General Relativity from the Fierz-Pauli action in Minkowski spacetime
by adding gravitational self-coupling terms consistent with diffeomorphism invariance.
On closer inspection, this process involves fixing boundary terms and different metric theories
can consequently be reconstructed~\cite{rubio}. 
Moreover, in order to account for large matter sources from the onset, the bootstrapped Newtonian
approach starts directly from a Fierz-Pauli-type action for the static Newtonian potential, to which
several terms are added, with coupling constants that vary from their
Einstein-Hilbert values in order to effectively accommodate for the (unknown) underlying
quantum dynamics.
The direct outcome of this programme is a non-linear equation that determines
the gravitational potential acting on test particles at rest, and which is generated by a static large source,
including pressure effects and the gravitational self-interaction to next-to-leading
order in Newton's constant.
\par
In this work, we will study solutions of the bootstrapped equation for the gravitational potential
generated by a localised source in $(1+1)$- and $(1+2)$-dimensional spacetimes.
The common feature that emerges is that the potential in the vacuum outside the compact
source has finite support and the outer space is therefore limited in size.
This can be interpreted as the necessary presence of a particle, or cosmological,
horizon associated with the compact source.
This result sets the $(1+1)$ and $(1+2)$-dimensional cases apart from the higher dimensional ones,
in which one can instead find solutions in the vacuum outside the compact source that asymptote
to zero at infinite distance.
We should also recall that General Relativity in one and two spatial dimensions is rather
different from the standard $(1+3)$-dimensional case.
For instance, there is no Newtonian limit in $(1+2)$-dimensional spacetime,
whereas this limit exists by construction in the bootstrapped Newtonian case.
The analysis in the present work is therefore also aimed at further clarifying the difference
between General Relativity and our non-linear version of Newtonian gravity.
\par
In line with the above-mentioned idea of dimensional reduction, we conjecture that these
lower-dimensional cases could describe finite size configurations accessible by high-energy
processes in four spacetime dimensions. 
What we have in mind here are processes that take place near the Planck scale, such as particle
scatterings at these energies, the gravitational collapse of compact objects forming black holes,
or the end stage of black hole evaporation. We expand on these ideas in the Conclusions.
\par
Moreover, we will see that the size of such ``baby universes'' turns out to depend on the
ADM mass~\cite{adm} and the value of the effective gravitational coupling.
Although the exact value of the lower (and higher) dimensional gravitational constant is unknown,
several generic forms for the $(1+d)$-dimensional one have been proposed. 
One such expression derived from entropic principles is given by~\cite{rmjmentropic}~\footnote{We will
set $c=1$ throughout the paper and usually denote $G_{(3)}=\gn=\lp/\mpl$, with $\lp$ and $\mpl$
the four-dimensional Planck length and mass, respectively.
Therefore, the Planck constant $\hbar=\lp\,\mpl$ has dimensions of length times mass and
$G_{(d)}$ of inverse mass times length$^{d-2}$.}
\be
G_{(d)}
=
2\,\pi^{1-\frac{d}{2}} \,\Gamma\left(\frac{d}{2}\right) \frac{\ell_{\rm p}^{d-1}}{\hbar}
=
2\,\pi^{1-\frac{d}{2}} \,\Gamma\left(\frac{d}{2}\right) \frac{\ell_{\rm p}^{d-2}}{\mpl}
\ ,
\label{gnval}
\ee
in which the presence of $\hbar$ is reminiscent of its quantum nature.
Using these expressions, we estimate that the typical order of magnitude for the
size of the lower-dimensional ``baby universes'' is Planckian for similar ADM mass.
\par
The paper is organised as follows:
in Section~\ref{S:action} we review the equation for the bootstrapped Newtonian potential
in $d\ge 1$ spatial dimensions;
the cases for $d=1$ and $d=2$ are then analysed in detail in Sections~\ref{S:d1} and \ref{S:d2},
respectively (the outer vacuum in $d\ge 3$ is briefly discussed in Appendix~\ref{S:d>3});
finally, we draw our conclusions in Section~\ref{S:conc}.
\section{Bootstrapped gravitational potential in $d$ spatial dimensions}
\label{S:action}
\setcounter{equation}{0}
We start by briefly reviewing the bootstrapped gravitational potential of
Refs.~\cite{Casadio:2019cux,Casadio:2019pli}, which we want here to
extend to the case of $d$ spatial dimensions.
This potential $V=V(r)$ is supposed to describe the gravitational pull
on test particles generated by a spherically symmetric and static matter density
$\rho=\rho(r)$ and is obtained from the Newtonian Lagrangian
\be
L_{\rm N}[V]
=
\Omega_{(d)}
\int_0^\infty
r^{d-1}\,\d r
\left(
\frac{(V')^2}{2\,\Omega_{(d)}\,G_{(d)}}
-\rho\,V
\right)
\ ,
\label{LagrNewt}
\ee
where $V'\equiv \d V/ \d r,$ $\Omega_{(d)}$ is the surface area of a $d$-dimensional unit sphere, given by
\be
\Omega_{(d)}
=
\frac{2\,\pi^{d/2}}{\Gamma\left(\frac{d}{2}\right)}
\ ,
\ee
and $G_{(d)}$ is Newton's constant in $d$ spatial dimensions.
The corresponding equation of motion is the Poisson equation for the $d$-dimensional
Newtonian potential $V=V_{\rm N}$, {\sl i.e.}
\be
r^{1-d}\left(r^{d-1}\,V'\right)'
\equiv
\triangle_{(d)} V
=
\Omega_{(d)}\,G_{(d)}\,\rho
\ .
\label{EOMn}
\ee
We can then include the effects of gravitational self-interaction by noting that the Hamiltonian
$H_{\rm N}[V]=-L_{\rm N}[V]$, computed on-shell by means of Eq.~\eqref{EOMn},
yields the Newtonian potential energy
\be
U_{\rm N}(r)
&\!\!=\!\!&
\frac{\Omega_{(d)}}{2}
\int_0^r 
{\bar r}^{d-1}\,\d {\bar r}
\,\rho(\bar r)\, V(\bar r)
\nonumber
\\
&\!\!=\!\!&
-\frac{1}{2\,G_{(d)}}\,
\int_0^r 
{\bar r}^{d-1} \,\d {\bar r}\,
\left[V'(\bar r) \right]^2
\ ,
\label{Unn}
\ee
where the boundary terms can be discarded through integration by parts 
(see Appendix~A in Ref.~\cite{Casadio:2019cux}).
One can view $U_{\rm N}$ as representing the interaction
of the matter distribution enclosed in a sphere of radius $r$ with the gravitational field.
Following Ref.~\cite{Casadio:2016zpl} (see also Ref.~\cite{dadhich1}),
we then define a self-gravitating source proportional to the gravitational
energy $U_{\rm N}$ per unit volume, that is 
\be
J_V
\simeq
\frac{\d U_{\rm N}}{\d \mathcal{V}} 
=
-\frac{2\left[ V'(r) \right]^2}{\Omega_{(d)}\,G_{(d)}}
\ .
\label{JV}
\ee
In Refs.~\cite{Casadio:2017cdv,BootN,Casadio:2019cux} we also included the source term 
\be
J_\rho=-2\,V^2
\ ,
\ee
which comes from the linearisation of the volume measure around the vacuum~\cite{Casadio:2017cdv}
and couples with the matter source according to General Relativity.
The corresponding interaction term can be viewed as a gravitational one-loop correction
to the matter density.
In Ref.~\cite{BootN}, we found that the pressure $p$ becomes very large for sources
with large compactness $R\sim \gn\,M$,
and we must therefore add a corresponding potential energy $U_{p}$ such that
\be
p
=
-\frac{\d U_{p}}{\d \mathcal{V}} 
\ .
\label{JP}
\ee
Since the latter contribution just adds to $\rho$, it can be easily included by simply shifting
$\rho \to \rho+q_p\,p$. Following this we obtain
\be
L[V]
=
-\Omega_{(d)}
\int_0^\infty
r^{d-1}\,\d r
\left[
\frac{\left(1-4\,q_V\,V\right)\left(V'\right)^2}{2\,\Omega_{(d)}\,G_{(d)}}
+V
\left(1-2\,q_\rho\,V\right)
\left(\rho+q_p\,p\right) 
\right]
\ ,
\label{LagrV}
\ee
where the non-negative coefficients $q_V$, $q_\rho$ and $q_p$ play the role of coupling constants
for the graviton current $J_V$, the matter current $J_\rho$ and the pressure $p$.
The Euler-Lagrange equation for $V$ is then given by
\be
\triangle_{(d)} V
=
\Omega_{(d)}\,G_{(d)}
\frac{1-4\,q_\rho\,V}{1-4\,q_V\,V}
\left(\rho+q_p\,p\right)
+
\frac{2\,q_V\left(V'\right)^2}
{1-4\,q_V\,V}
\ .
\label{EOMV}
\ee
For simplicity, we set $q_V=q_\rho=q_p$ from here on.~\footnote{
As we will see in the following Sections, the main results in the present work consider only the potential
in the region outside localised sources, where the values of $q_\rho$ and $q_p$ do not matter.}
Finally, the conservation equation that determines the pressure reads
\be
p'
=
-V'\left(\rho+p\right)
\ .
\label{eqP}
\ee
\par
In all spatial dimensions, the mass is simply modelled as a spherically-uniform density
distribution of radius $R$,
\be
\rho
=
\rho_0
\equiv
\frac{M_0}{{\cal V}_{(d)}(R)}\, 
\Theta(R-r)
\ ,
\label{HomDens}
\ee
where $\Theta$ is the Heaviside step function, and
\be
{\cal V}_{(d)}(R) = \frac{\pi^\frac{d}{2}\,R^d}{\Gamma\left(1+{d}/{2}\right)}
\ee
is the volume of a $d$-dimensional sphere. The proper mass is given by
\be
M_0
=
\Omega_{(d)}
\int_0^R
r^{d-1}\,\d r\,\rho(r)
\ .
\ee
Regularity conditions in the centre are required to be met by the solutions, specifically
\be
V_{\rm in}'(0)=0
\label{b0}
\ee
and they must also satisfy matching conditions with the exterior solution at the surface, 
\be
V_{\rm in}(R)
&\!\!=\!\!&
V_{\rm out}(R)
\equiv
V_R
\label{bR}
\\
\notag
\\
V'_{\rm in}(R)
&\!\!=\!\!&
V'_{\rm out}(R)
\equiv 
V'_R
\ ,
\label{dbR}
\ee
where we defined $V_{\rm in}=V(0\le r\le R)$ and $V_{\rm out}=V(R\le r)$.
These conditions allow us to solve for the pressure in Eq.~\eqref{eqP} in terms
of the potential, which yields the general differential equation
\be
\triangle_{(d)} V
=
\frac{\Omega_{(d)}\,G_{(d)}\,M_0}{{\cal V}_{(d)}(R)}\, e^{V_{R} - V}
+ 
\frac{2\,q_V\,(V^\prime)^2}{1-4\,q_V\,V}
\ .
\label{ints}
\ee
\par
In the vacuum ($\rho=p=0$), Eq.~\eqref{eqP} is trivially satisfied and Eq.~\eqref{EOMV}
reads
\be
\frac{\left(r^{d-1}\,V'\right)'}
{r^{d-1}}
=
\frac{2\,q_V \left(V'\right)^2}{1-4\,q_V\,V}
\ .
\label{EOMV0}
\ee
For $q_V>0$, we can rescale $\tilde V=q_V\,V$  so that Eq.~\eqref{EOMV0}
can be written
\be
\frac{\left(r^{d-1}\,\tilde V'\right)'}
{r^{d-1}}
=
\frac{2 \left(\tilde V'\right)^2}{1-4\,\tilde V}
\ ,
\label{tEOMV0}
\ee
where it becomes apparent that $\tilde V$ does not depend on $q_V$ but only on $d$.
\par
In the following sections, we will consider in detail the cases $d=1$ and $d=2$,
for which we also briefly reference the geometric pictures for comparison.
\section{One-dimensional case}
\label{S:d1}
\setcounter{equation}{0}
An early study of $(1+1)$-dimensional Schwarzschild-like black holes found that a proper Newtonian
limit could be obtained, giving the expected equations of motion~\cite{rbm11}.
From the action
\be
S
=
-\int \d^2x\,\sqrt{|g|}\, \psi \left(\frac{R -\Lambda}{8\,\pi\, G_{(1)}} - {\cal L}_{\rm m}\right)
\ ,
\ee
the Euler-Lagrange equations yield a metric that is independent of the dilaton field $\psi$.
The matter Lagrangian is directly related to the stress-energy tensor via its
trace ${\cal L}_{\rm m} = T^\alpha_{~\alpha} = T$. 
Here, $G_{(1)}$ is the one-dimensional gravitational constant~\footnote{Note that $G_{(1)}$ 
has dimensions of the inverse of mass times length, therefore $G_{(1)}\,\hbar$ is dimensionless.}
and we included a cosmological constant $\Lambda$ for a reason that
will become apparent shortly. 
\par
For a point-like source of mass $M$ located at $x=x_0$,
\be
\rho = M\,\delta(x-x_0)
\ ,
\label{Sd1}
\ee
the metric reads~\cite{rbm11}
\be
\d s^2
=
\left(1-2\,G_{(1)}\,M\,r+r^2/\ell^2\right) \d t^2
- \frac{\d x^2}{1-2\,G_{(1)}\,M\,r+r^2/\ell^2}
\ ,
\label{11metric}
\ee
where $r\equiv |x-x_0|$ and we assumed $\ell=\sqrt{2/\Lambda}>0$.
Note that this metric depends only on the relative coordinate $r$ with respect
to the location of the source, and becomes linear in $r$ in the limit $\ell\to\infty$
(equivalently, $\Lambda\to 0$).
In this limit, one also finds that the one-dimensional version of the Newtonian
potential is linear,
\be
V_{\rm N} = 
1-g_{tt}/2
=
G_{(1)} \,M\,r
\ .
\label{Vn1}
\ee
The equations of motion of a test particle can be derived from the geodesic equation,
and reveal the expected $(1+1)$-dimensional kinematic equations.
\par
Note that for sufficiently large $\ell$, the metric~(\ref{11metric}) admits both a black hole
horizon with size
\be
\Rh
=
\ell
\left(
\ell\,G_{(1)}\,M
-\sqrt{\ell^2\,G_{(1)}^2\,M^2-1}
\right)
\label{rh1-1}
\ee
and a cosmological horizon with size
\be
R_\Lambda
=
\ell
\left(
\ell\,G_{(1)}\,M
+\sqrt{\ell^2\,G_{(1)}^2\,M^2-1}
\right)
\simeq
{2\,\ell^2\,G_{(1)}\, M}
\ .
\label{rL1-1}
\ee
In the limit $\ell\gg (G_{(1)}\,M)^{-1}$, one obtains
\be
\Rh
\simeq
\frac{1}{2\,G_{(1)}\, M}
\ ,
\label{11horizonH}
\ee
which scales as the inverse of the mass,
and
\be
R_\Lambda
\simeq
{2\,\ell^2\,G_{(1)}\, M}
=
\frac{\ell^2}{\Rh}
\ .
\label{11horizonL}
\ee
Comparing this to the general expression for $(1+d)$-dimensional black holes,
\be
\Rh
=
\left(2\,G_{(d)}\,M\right)^\frac{1}{d-2}
\ ,
\label{genhorizon}
\ee
it can be seen that the result~(\ref{11horizonH}) in the limit $\ell\to\infty$
can be recovered with the naive substitution $d=1$.
This is also true for the generalized Newtonian potential
\be
V_{\rm N} = -\frac{G_{(d)}\,M}{r^{d-2}}
\ ,
\ee
except that the overall sign must be reversed for $d=1$.
In this sense, the $(1+1)$-dimensional spacetime follows the same general pattern
as do the quantities in higher dimensions.
\subsection{Bootstrapped vacuum}
Equation~\eqref{tEOMV0} in one spatial dimension ($d=1$) yields the solution
\be
V
=
\frac{\tilde V}{4\,q_V}
=
\frac{1}{4\,q_V}
\left[
1-
\left(\tilde C_1-\tilde C_2\,x\right)^{2/3}
\right]
\ ,
\label{V1ccc}
\ee
which does not contain an exact linear term.
In order to compare this potential with the classical geometry described above, we 
need to relate the two integration constants $\tilde C_1$ and $\tilde C_2$
to $M$ and $\ell$ (or, equivalently, $\Rh$ and $R_\Lambda$).
This can be done by requiring that the limit $q_V\to 0$ reproduces the Newtonian
potential obtained from the geometry.
\par
We first notice that the limit $q_V\to 0$ yields a finite potential only provided $\tilde C_2=q_V/L$
and $\tilde C_1=1+q_V\,x_0/L$, where $L$ and $x_0$ are constants to be determined.
In this case, we have
\be
V
=
\frac{1}{4\,q_V}
\left[
1-
\left(1-q_V\,\frac{x-x_0}{L}\right)^{2/3}
\right]
\ ,
\label{V1cc}
\ee
and we can identify $x_0$ with the location of the point-like source~\eqref{Sd1},
so that 
\be
V'(x_0)
=
\frac{1}{6\,L}
=
G_{(1)}\,M
=
V_{\rm N}'(r=0)
\ ,
\ee
where we used Eq.~\eqref{Vn1} for the expected Newtonian behaviour in the limit $q_V\to 0$.
The above relation uniquely determines 
\be
L^{-1}
=
6\,G_{(1)}\,M
\ ,
\ee
so that 
we finally obtain
\be
V
=
\frac{1}{4\,q_V}
\left[
1-
\left(1
-
6\,q_V\,G_{(1)}\,M\,r
\right)^{2/3}
\right]
=
\frac{1}{4\,q_V}
\left[
1-
\left(1
-
6\,q_V\,X\,\frac{r}{R}
\right)^{2/3}
\right]
\ ,
\label{V1out}
\ee
where the dimensionless quantity playing the role of the ``outer'' compactness in this case is given by
$X=G_{(1)}\,M\,R$, and $r\equiv x-x_0>0$.~\footnote{The mirror region $x-x_0<0$ can be obtained
by replacing $L\to-L$ in Eq.~\eqref{V1cc}.}
\par
Since $V(0)=0$, the black hole horizon can be obtained from the Newtonian
condition $2\,V(\rh)=1$, or
\be
\rh
=
\frac{1-(1-2\,q_V)^{3/2}}{6\,q_V\,G_{(1)}\,M}
\ ,
\label{11horizon}
\ee
so that, for $0<q_V\ll 1$, we have
\be
\rh
\simeq
\Rh
-
q_V\,\Rh/2
\ ,
\label{rhq0}
\ee
and the expected geometric result~\eqref{11horizonH} is precisely recovered for $q_V\to 0$.
Unlike in $d=3$, however, the black hole horizon exists only for $q_V<1/2$. 
\par
\begin{figure}[t]
\centering
\includegraphics[width=10cm]{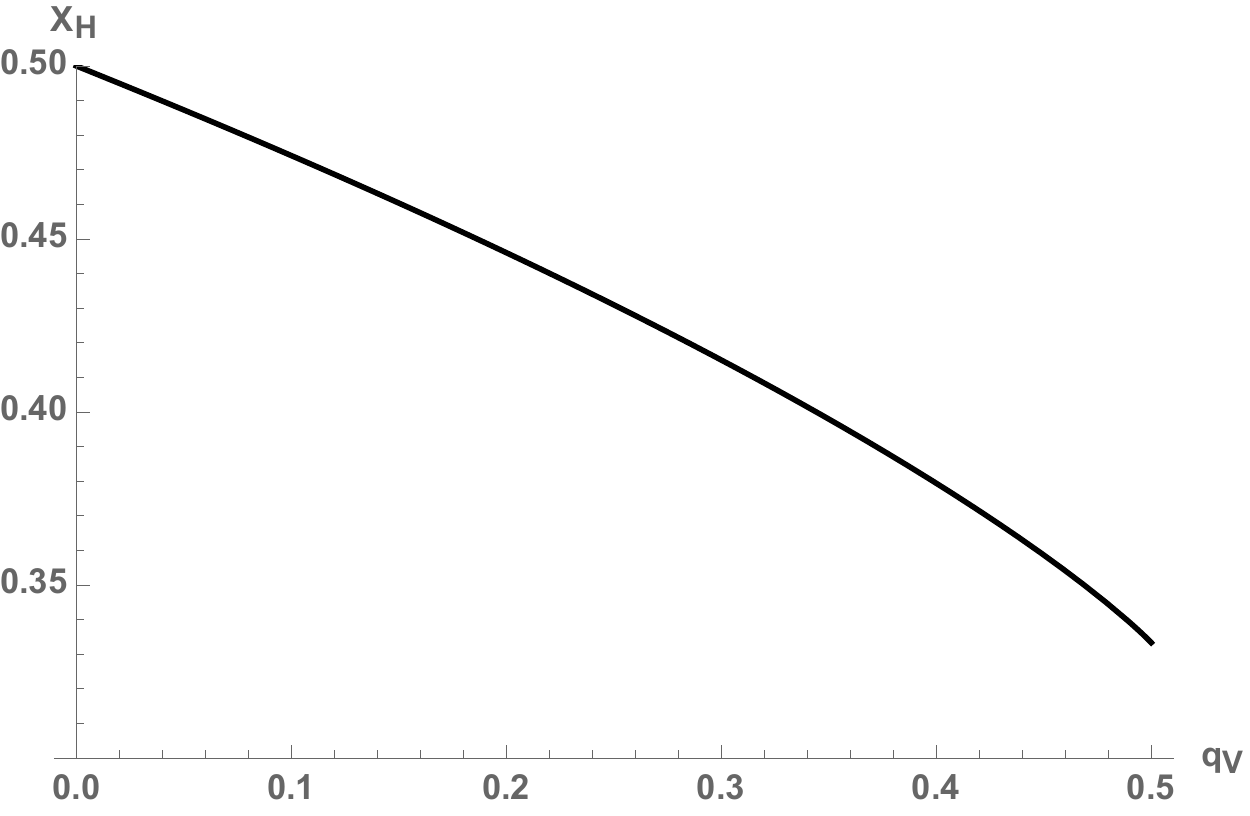}
\caption{Compactness $X_{\rm H}$ as a function of $q_V$.}
\label{X_H_1d}
\end{figure}

The compactness value for which the size of the object equals the horizon size is 
\be
X_{\rm H}
=
\frac{1-(1-2\,q_V)^{3/2}}{6\,q_V}
\ ,
\ee
and, as expected only depends on the coupling $q_V$. A plot showing this dependence
is presented in Fig.~\ref{X_H_1d}.
\par
In fact, the solution~\eqref{V1cc} exists only for 
\be
0<r
\le
L/q_V
\ .
\ee
By assuming the upper limit is the cosmological horizon,
\be
R_\Lambda
=
\frac{L}{q_V}
=
\frac{1}{6\,q_V\,G_{(1)}\,M}
\ ,
\label{cosmoh}
\ee
we can identify a cosmological constant as
\be
\Lambda
=
24\,q_V\,G_{(1)}^2\,M^2
\ .
\ee
The radius of this cosmological horizon diverges in the limit $q_V\to 0$ for finite $M$,
which is also in agreement with the result in Eq.~\eqref{rhq0}.
\par
\begin{figure}[t]
\centering
\includegraphics[width=10cm]{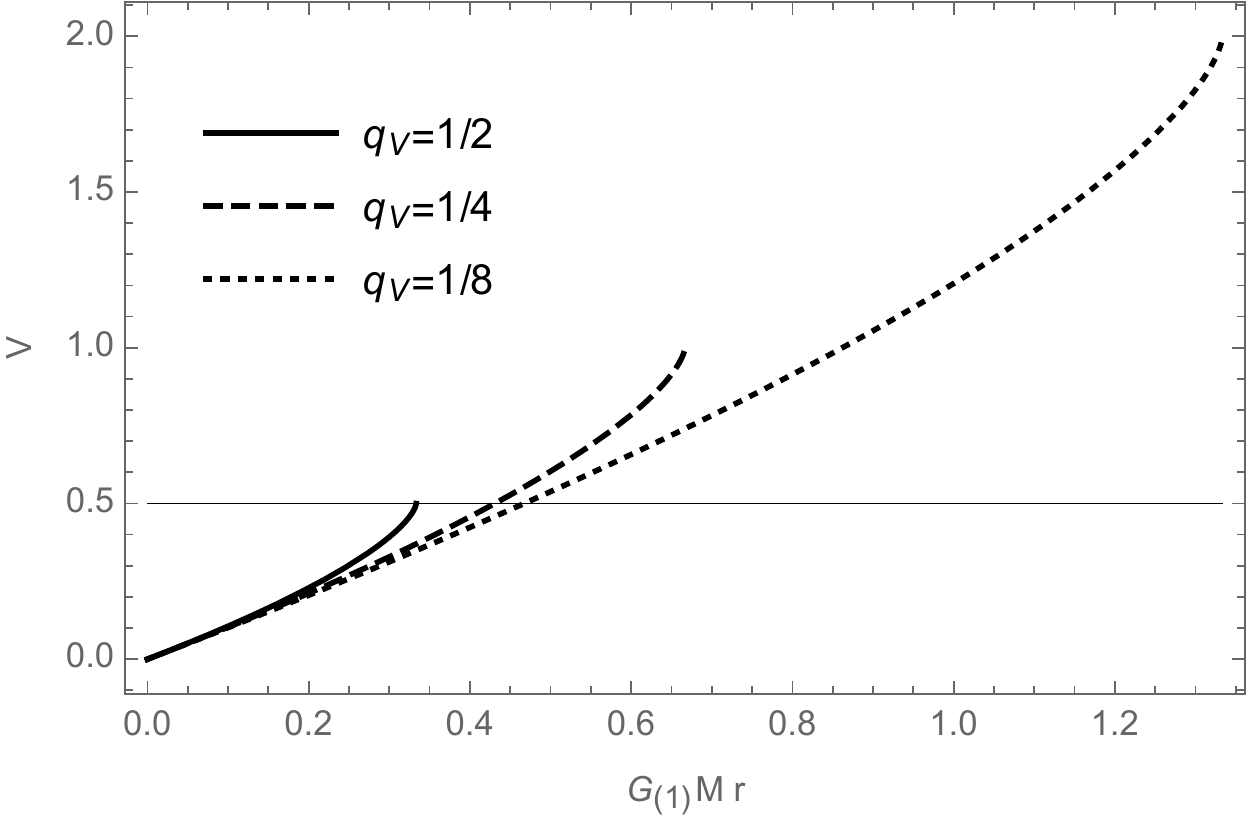}
\caption{Potential $V$ in $d=1$.
The intersections with the thin horizontal line mark the locations of the
black hole horizons.}
\label{pV1}
\end{figure}

A curious feature is that the black hole horizon and the 
cosmological horizon scale in the same way with the mass $M$, and their ratio 
\be
\frac{\Rh}{R_\Lambda}=1-(1-2\,q_V)^{3/2}
\ee
only depends on the coupling $q_V$.
This behaviour differs significantly from the classical case described by Eqs.~\eqref{rh1-1}
and~\eqref{rL1-1}, in that the cosmological constant $\Lambda$ and the mass $M$ cannot
be varied independently.
That is, whenever there is a localised source $M$, there is also a cosmological horizon determined
by the gravitational self-coupling $q_V$.
Upon requiring $R_\Lambda>\rh$, we again obtain $q_V<1/2$,
which explains why there cannot be a black hole horizon for $q_V>1/2$.
A few examples are plotted in Fig.~\ref{pV1}. As expected from the above expression, the plots show that 
the black hole horizon coincides with the cosmological horizon
for $q_V=1/2$ (solid line). 
\par
Regarding the size $R_\Lambda$ of these ``universes'',  we first note that the general expression~\eqref{gnval}
for the gravitational coupling for $d=1$ reduces to
\be
G_{(1)}
=
\frac{2\,\pi}{\hbar}
\ ,
\ee
indicating that $G_{(1)}$ is truly a quantum constant and thus $(1+1)$-dimensional gravity is similarly quantum.
{\it Vis-\`a-vis\/} the cosmological horizon~\eqref{cosmoh}, one can estimate
\be
R_\Lambda
=
\frac{\hbar}{12\,\pi\,q_V\,M}
=
\frac{\lp}{12\,\pi\,q_V}\,
\frac{\mpl}{M}
\ .
\label{hcosmoh}
\ee
Since we must have $q_V<1/2$, this is about an order of magnitude short of $\lp$
for $M \sim \mpl$, and suggests that there can be ``baby Planck universes''
that nucleate around such objects.
\par
In the above analysis of the vacuum generated by matter,
we made use of a point-like source in order to fix
arbitrary coefficients and compare with the geometric description.
However, a regular source is part of the bootstrapped picture
and we are therefore going to consider that case next.
The purpose of the following analysis is merely to show the existence
of regular interior solutions without the pretence of being exhaustive.
For that reason we will simply consider uniform sources whose energy density
is given by Eq.~\eqref{HomDens} with $d=1$.
Any more realistic case would indeed require numerical methods beyond the scope
of this work.
\subsection{Bootstrapped interior}
\label{SS:bv1}
For $d=1$, Eq.~\eqref{ints} reads
\be
V''
=
\frac{G_{(1)}\,M_0}{R}\, e^{V_{R} - V}
+ 
\frac{2\,q_V\,(V^\prime)^2}{1-4\,q_V\,V}
\ .
\ee
In the small compactness regime where $X\lesssim X_{\rm H}$, we find the approximate solution
\be
V
\simeq
V_{0}+\frac{G_{(1)} \,M_{0}\, e^{V_{R} - V_{0}}}{2\,R} \,r^2
\ ,
\label{potential1d}
\ee 
and after requiring continuity between inner and outer solutions at $r=R$, we obtain
\be
2\,V_R
&\!\!\simeq\!\!&
2\,V_0
+
G_{(1)} \,M_{0}\, R \, e^{V_R-V_0}
\nonumber
\\
\label{match1}
\\
V_R^\prime
&\!\! \simeq\!\! &
G_{(1)}\,M_{0} \,e^{V_R-V_0}
\ ,
\nonumber
\ee
which gives
\be
V_0 
\simeq
\frac{1}{4\,q_V}\left[1-\left(1-6\,q_V\, X\right)^{2/3}\right]
+ \ln\left[\frac{M_{0}}{M} \left(1-6\,q_V\, X\right)^{1/3}\right]
\ .
\ee
The mass constraint yields
\be
M_{0}
\simeq
\frac{M\,e^{-\frac{X}{2 \left(1-6\,q_V\,X\right)^{1/3}}}}
{\left(1-6\,q_V\,X\right)^{1/3}}
\ ,
\ee
and so the approximate potential is found to be
\be
V
\simeq
 \frac{1}{4q_V}\left[1-\left(1-6\,q_V \, X\right)^{2/3}\right]
 -
 \frac{X}{2 \left(1-6\,q_V\,X\right)^{1/3}}
 \left(1-\frac{r^2}{R^2}\right)
 \ .
\ee
\par
\begin{figure}[t]
\centering
\includegraphics[width=10cm]{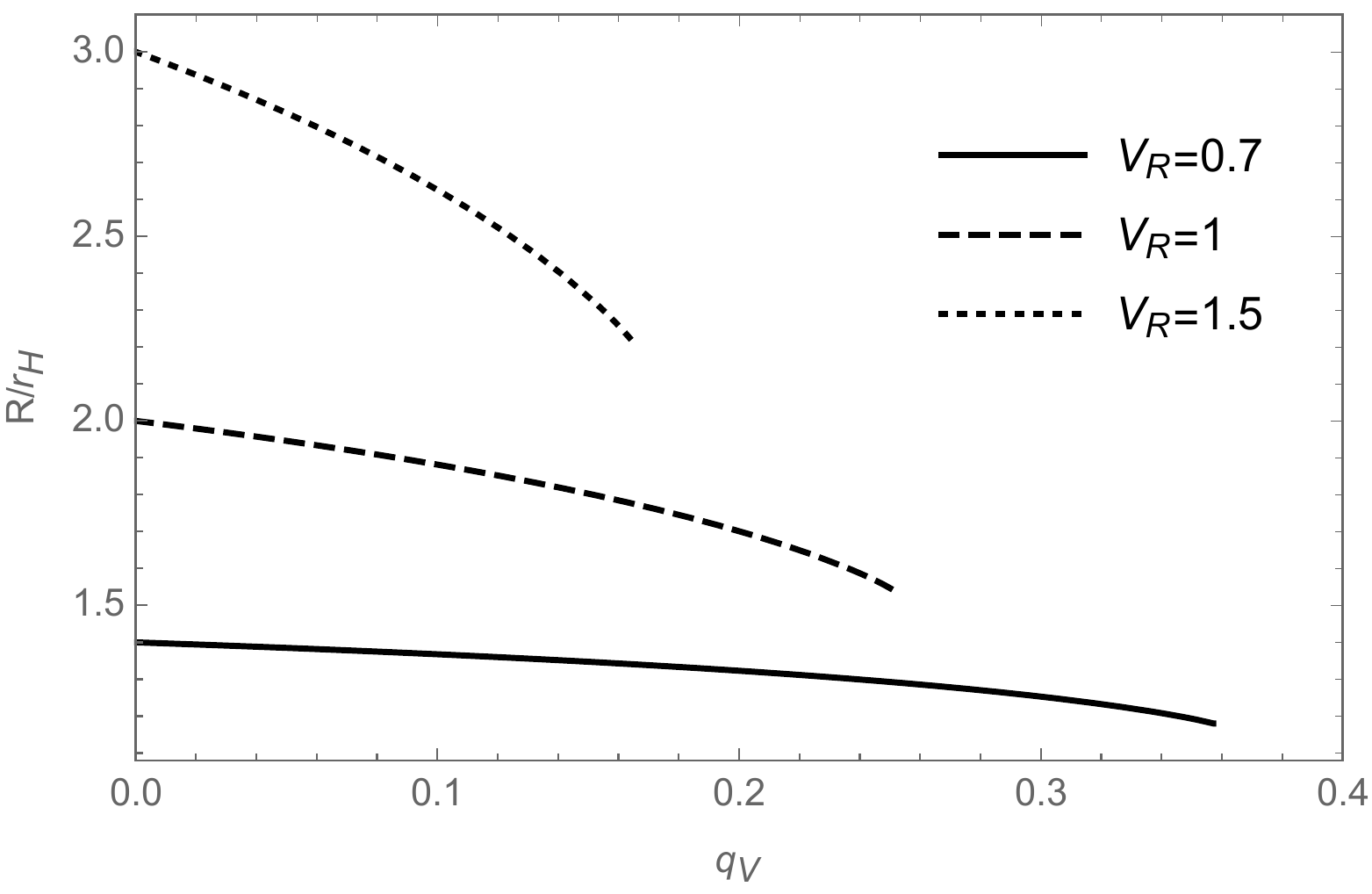}
\caption{Ratio (\ref{rrat}) of the size $R$ of a star with mass $M$ with surface potential $V_R$
to the horizon size $\rh$ of an equal-mass black hole.
The curves are for $V_R = 0.7$ (solid), $1.0$ (dashed), and $1.5$ (dotted).
The right-hand cutoff is for the value of $q_V$ determined by (\ref{qmax}).}
\label{RRatio}
\end{figure}

Since the Newtonian potential in $(1+1)$-dimensions grows linearly with distance,
one expects non-compact objects (stars) have a surface potential that satisfies $2\,V_R> 1$. 
For a given $M$, the above expression can be solved for $R$ to give
\be
R
\simeq
\frac{1-\left(1-4\,q_V\, V_R\right)^{3/2}}
{6\,q_V \,G_{(1)}\,M}
\ .
\ee
This imposes a constraint on $V_R$ that depends on $q_V$, namely
\be
V_R
\leq
\frac{1}{4\,q_V}
\ .
\label{qmax}
\ee
Comparing this expression to Eq.~\eqref{11horizon}, we find
\be
\frac{R}{\rh}
\simeq
\frac{1-\left(1-4\,q_V\, V_R\right)^{3/2}}
{1-\left(1-2\,q_V\right)^{3/2}}
\ .
\label{rrat}
\ee
In the limit $2\,V_R = 1$ ({\it i.e.}~a black hole), the ratio is unity. 
It can be seen from the above expression, however, that $R>\rh$
for any $V_R > {1}/{2}$.
This is displayed in Figure~\ref{RRatio}.
\section{Two-dimensional case}
\label{S:d2}
\setcounter{equation}{0}
General Relativity in $(1+2)$-dimensions is quite non-intuitive.
The best known solution is the BTZ metric for a rotating black hole with angular momentum $J$ in a spacetime with
cosmological constant $\Lambda$~\cite{btz},
\be
\d s^2
&\!\!=\!\!&
\left(-G_{(2)}\,M + \Lambda\,r^2+\frac{J^2}{4\,r^2}\right)^{1/2}\d t^2
- \frac{\d r^2}{\left(-G_{(2)}\, M + \Lambda\, r^2+\frac{J^2}{4\,r^2}\right)^{1/2}}
\nonumber
\\
&&
- r^2 \left(\d\phi + \frac{J^2}{2\,r^2}\,\d t\right)^2
\ .
 \ee
In this case, the gravitational  constant $G_{(2)}$ has dimensions of inverse mass.
When $J=0$ ({\it i.e.} static black holes), it is clear that horizons exist only for negative $\Lambda$,
\be
R_{(2)}
=
\sqrt{-\frac{G_{(2)}\,M}{\Lambda}}
\ .
\ee
That is, $(1+2)$-dimensional black holes can only exist in AdS spacetimes. 
Furthermore, there is no Newtonian limit in this case and gravity is purely topological.
Note also that the form of the horizon does not follow the pattern established in
Eq.~\eqref{genhorizon} when $d=2$ since the expression is not defined at this point.
\subsection{Bootstrapped vacuum}
In two spatial dimensions, the solution to Eq.~\eqref{tEOMV0} is

\be
V
=
\frac{\tilde V}{4\,q_V}
=
\frac{1}{4\,q_V}
\left\{
1-
\left[C_1-C_2\,\ln( r/\ell)\right]^{2/3}
\right\}
\ ,
\ee
where $\ell$ is an arbitrary length scale.
The expected Newtonian behaviour is given by
\be
V_{\rm N}
=
G_{(2)}\,M\,\ln(r/\ell)
\ .
\label{Vn2}
\ee
If we set $C_1=1$ and 
\be
C_2=6\,q_V\,G_{(2)}\,M
\ ,
\ee
we thus obtain
\be
V
=
\frac{1}{4\,q_V}
\left\{
1-
\left[1-6\,q_V\,G_{(2)}\,M\,\ln( r/\ell)\right]^{2/3}
\right\}
\ .
\label{V2out}
\ee
In the limit $\ln(r/\ell)\ll 1$, {\it i.e.} for $r\simeq\ell$, this can be written as
\be
V
\simeq
G_{(2)}\,M\,\ln(r/\ell)
+
q_V\,G_{(2)}^2\,M^2\left[\ln(r/\ell)\right]^2
\ .
\ee
\par
We can find the location of the black hole horizon by imposing the Newtonian
condition $2\,V(\rh)=-1$, which yields
\be
\rh
\simeq
\ell\,\exp\left[\frac{1-(1+2\,q_V)^{3/2}}{6\,q_V\,G_{(2)}\,M}\right]
\ .
\label{rh2}
\ee
\par
\begin{figure}
\centering
\includegraphics[width=10cm]{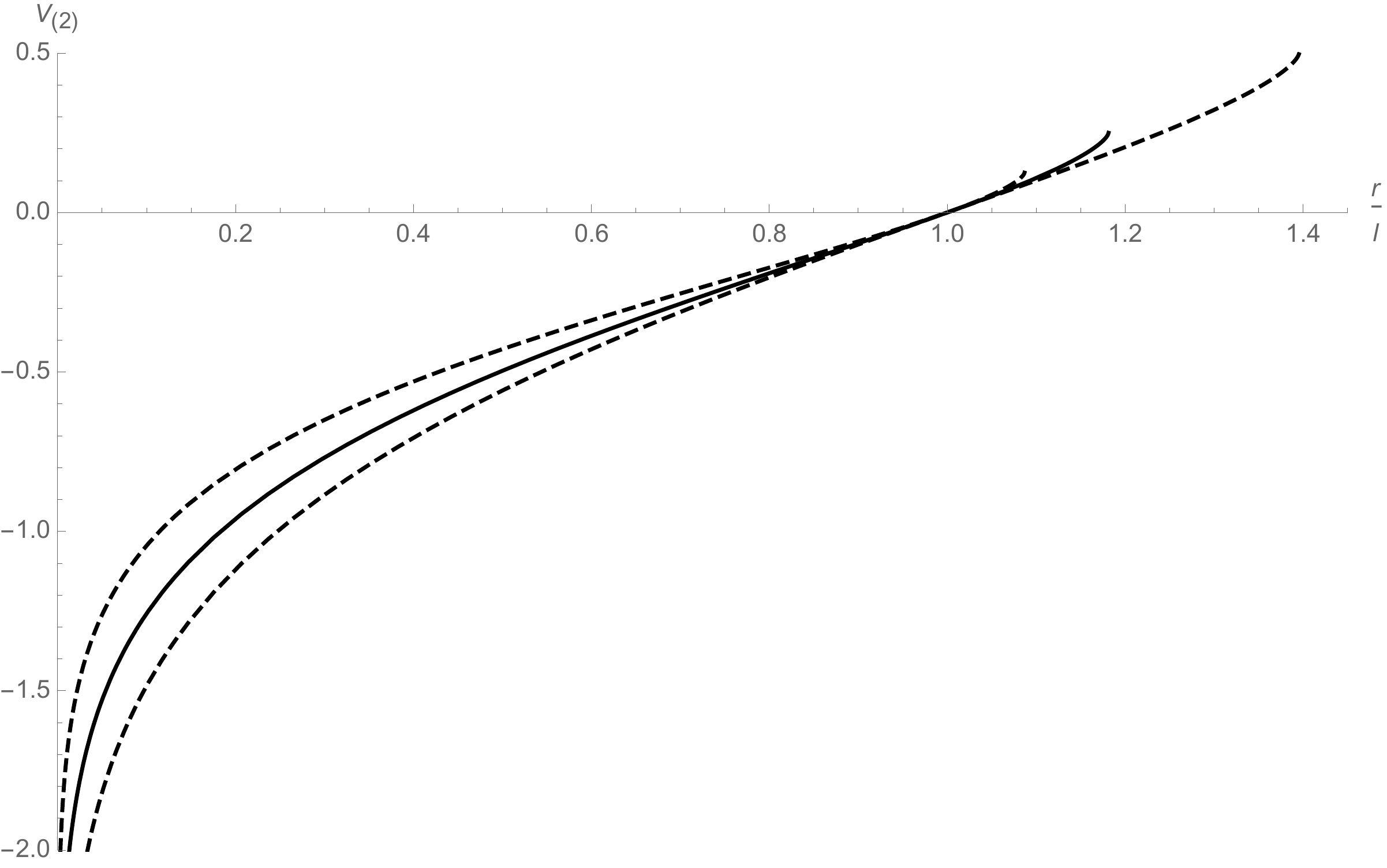}
\caption{Potential $V_{(2)}$ for $q_V=1/2$ (dotted line), $q_V=1$ (solid line)
and $q_V=2$ (dashed line).}
\label{pV2}
\end{figure}

Note that for positive values of the self-interaction coupling $q_V$ and mass $M$, the
solution~\eqref{V2out} only exists for 
\be
0< \frac{r}{\ell} \le \exp\left(\frac{1}{6\,q_V\,G_{(2)}\,M}\right)
\ , 
\label{rconstraint}
\ee
which means that a cosmological horizon can be assumed to be present at 
\be
R_\Lambda
=
\ell\, \exp\left(\frac{1}{6\,q_V\,G_{(2)}\,M}\right)
\ .
\label{cosmoh2}
\ee
In analogy with the $(1+1)$-dimensional case, and to emphasise the difference with
the classical solutions, we note that the ratio between the black hole and cosmological horizons is given by
\be
\frac{\rh}{R_\Lambda}
\simeq
\exp\left[-\frac{(1+2\,q_V)^{3/2}}{6\,q_V\,G_{(2)}\,M}\right]
\ ,
\ee
which is now completely determined by the mass $M$ and the coupling $q_V$.
A few examples for the potential are plotted in Fig.~\ref{pV2}.
\par
To estimate the size of the horizons, we again appeal to the expression~\eqref{gnval},
which now gives
\be
G_{(2)}
=
\frac{2\,\lp}{\hbar}
\sim
\frac{1}{\mpl}
\ .
\label{g2}
\ee
The cosmological horizon will then scale as
\be
R_\Lambda
\sim
\ell\,\exp\left(\frac{\mpl}{6\,q_V\,M}\right)
\ .
\ee
Unlike in $d=1$, this result depends also on the arbitrary scale $\ell$.
Thus, the value of $\ell$ can be constrained by placing an upper limit on the possible size of such horizons.
If these are taken to be roughly Planck scale as well, then $\ell \sim \lp$ is a reasonable assumption. 
\subsection{Bootstrapped interior}
Repeating the analysis of Section~\ref{S:d1}, we are now going to show that mathematically
regular solutions for the interior of the source can also be found for $d=1$.
The form of the interior solution depends on the compactness of the source, 
which is now given by
\be
X
=
G_{(2)}\,M
\ ,
\ee
so the derivation must be handled in different ways depending on the relative size of this quantity.  
For small compactness, $X\ll 1$, the differential equation (\ref{ints}) can be solved as a series and we find to order $r^2$
\be          
V(r)
\simeq
V_{0} + \frac{G_{(2)}\,M_0\, e^{V_{R} - V_{0}}}{2\,R^2}\, r^2
\ ,
\label{potential2d}
\ee
The form of the above potential is identical to the standard $(1+3)$-dimensional case, up to the power of $R$
in the denominator of the second order term. 
That is, the dependence on $r$ is consistently $r^2$.  
\par
The matching conditions~\eqref{bR} and \eqref{dbR} imply
\be
2\left(V_R-V_0\right)
&\!\!\simeq\!\!&
G_{(2)}\,M_{0}\, e^{V_R-V_0}
\nonumber
\\
\label{match2}
\\
R\, V'_R
&\!\! \simeq\!\! &
G_{(2)}\, M_{0}\, e^{V_R-V_0}
\ .
\nonumber
\ee
The equation that ensures continuity of the first derivatives of the potentials across the boundaries
in the two cases can be used to first obtain
\be
V_0
\simeq
\frac{1}{4\,q_V}
\left\{1-
\left[1-6\,q_V\, X \ln\left(\frac{R}{\ell}\right)\right]^{2/3}
\right\}
+
\ln\left\{\frac{M_{0}}{M} \left[1-6\,q_V\, X\ln\left(\frac{R}{\ell}\right)\right]^{1/3}
\right\}
\ .
\ee
Again using the equations which ensure the continuity of the potentials, one can write $M_{0}$ in terms of $M$ as
\be
M_{0}
\simeq
\frac{M \, e^{-\frac{X}{2 \left[1-6\,q_V\, X\ln\left(R/\ell\right)\right]^{1/3}}}}
{\left[1-6 \,q_V\,X\ln\left(R/\ell\right)\right]^{1/3}}
\ .
\ee
This allows one to find the approximate inner potential
\be
V(r)
\simeq
\frac{1}{4\,q_V}
\left\{
1-\left[1-6\,q_V\, X \ln\left(\frac{R}{\ell}\right)\right]^{2/3}
\right\}
-
\frac{X}{2 \left[1-6\,q_V\, X\ln\left(R/\ell\right)\right]^{1/3}}
\left(1-\frac{r^2}{R^2}\right)
\ .
\ee
\par
As with the $(1+1)$-dimensional interior solution, we can solve the above potential
at the surface $r=R$ for a given mass $M$, which gives
\be
R
\simeq
\ell\,
\exp\left[\frac{1+\left(1-4\,q_V\, V_R\right)^{3/2}}{6\,q_V\, G_{(2)}\,M}\right]
\ ,
\ee
which reduces to the horizon~\eqref{rh2} when $2\,V_R = -1$.
The ratio of the two is 
\be
\frac{R}{\rh}
\simeq
\exp\left[\frac{(1-4\,q_V V_R)^{3/2}-(1+2\,q_V)^{3/2}}{6\,q_V\, G_{(2)}\,M}\right]
\ee
and so for $V_R < -1/2$ the black hole horizon will be smaller than $R$ for a given mass.  
\section{Conclusions}
\label{S:conc}
\setcounter{equation}{0}
We have studied the bootstrapped potential generated by compact sources in $d=1$ and
$d=2$ spatial dimensions and compared the associated characteristics with more standard descriptions.
It appears that the lower-dimensional cases remain at odds with those at $d\ge 3$,
perhaps even more so in the bootstrapped description of gravity.
In particular, the emergence of a compact spacetime associated to these sources
is intriguing, and we interpret these lower-dimensional configurations
as ``bubbles'' possibly generated by ultra high-energy processes in our $d=3$
Universe.
Examples of such processes could be particle scatterings at Planckian energies,
the gravitational collapse of compact objects leading to unbounded energy densities
near the centre of mass, or the end stage of black hole evaporation according to Hawking's
prediction that the black hole temperature should increase steadily as the mass decreases.
Although several parameters are involved, the size of these ``baby universes'' can
naturally be on the order of the Planck length for similarly scaled masses. 
\par
Furthermore, the bootstrapped Newtonian potentials for $d=1$ and $d=2$ 
contain a ``cosmological horizon'' at $r=R_\Lambda$ given respectively in Eq.~\eqref{cosmoh}
and Eq.~\eqref{cosmoh2}.
The size of the horizon~\eqref{cosmoh}, however, is solely determined by the localised mass $M$
in $d=1$, whereas the cosmological horizon~\eqref{cosmoh2} in $d=2$ is independent of $M$.
The origin of this difference is the appearance of an arbitrary length scale $\ell$
in the logarithm of the Newtonian limiting solution~\eqref{Vn2}.
One should then further recall that in $d\ge 3$ dimensions the ``cosmological horizon'' does not
need to be present at all, since the bootstrapped potential is defined for all values
of $r>0$.~\footnote{It is nonetheless interesting to investigate how an effective cosmological 
constant would appear in bootstrapped Newtonian gravity.
Preliminary results can be found in Refs.~\cite{cosmo} and a more detailed analysis is being
performed.}
Moreover, Newton's constant in $d=1$ equals the inverse of Planck's and the radius $R_\Lambda$
therefore entails a quantum nature in one spatial
dimension, clearly displayed in Eq.~\eqref{hcosmoh}.
\par
One of the interesting consequences of these baby universes is that they would replicate
the process of dimensional reduction, in which the dimensionality of spacetime decreases
as one approaches higher energies (equivalently, smaller length scales)~\cite{thooft,dsreview,dimreg} 
and the system becomes more and more quantum in nature.
This is a shared feature of many different approaches to quantum gravity. 
Here, the $d=1$ phase could be enveloped within the $d=2$ bubble,
particularly since the estimated scales in this work are within an order of magnitude of each other. 
\par
Finding a novel test of a theory whose fundamental energy scale is in the Planck range is problematic,
given today's technological limitations.
The highest energies that are possible to probe in laboratory experiments reach the TeV scale at best, 
as the recent LHC runs demonstrate.
Nature, however, has no such constraints and the available energies are much higher.
For example, ultrahigh energy cosmic rays have been measured in the $\sim 10^{20}~$eV range \cite{PAC}.
Although they are still far from the Planck scale, these processes mark a considerable improvement
over accelerator technology.
\par
The most promising window of opportunity to observe Planck scale physics, and by association
quantum gravity effects, currently rests in the realm of black holes.
With the detection of gravitational waves from compact binary coalescence by the LIGO/Virgo
collaboration~\cite{LIGOScientific:2018mvr} comes the possibility of identifying ``echoes'' in the
ring-down phase of the merger~\cite{echos}. 
This is predicated on the idea that there exists a near-horizon structure on the Planck scale that
generates periodic gravitational wave reflections which, if detected, would mark the first clear signal
of quantum gravity.
\par
Apart from gravitational waves, the advent of the Event Horizon Telescope and its triumphant imaging of the
supermassive black hole in M87~\cite{Goddi:2019ams} has introduced the possibility of studying near-horizon
physics on macroscopic scales in the electromagnetic spectrum.
Since quantum gravity effects are expected to become important in this regime, precision measurements
of the black hole shadow's size and morphology will yield crucial information about the nature of the spacetime
in that vicinity.
Planck scale metric fluctuations~\cite{Wheeler:1955zz} are expected to be amplified in such
a way as to produce measurable deviations from General Relativity~\cite{Giddings:2016tla}.
\par
The other avenue of testing quantum gravity is black hole evaporation.
Unlike the previous cases, these effects emerge only when the black hole horizon itself reaches
the Planck scale at the end of its life.
The standard Hawking picture is expected to break down at this stage, and quantum gravity effects
-- however they might manifest themselves -- will take over.
Searching for observational (electromagnetic or gravitational) evidence of the final stages of a
primordial black hole is thus important here.
Fast Radio Bursts (FRBs), for example, may be a result of the end stage of black hole evaporation,
which could again reveal fingerprints of the underlying quantum gravity theory~\cite{rovelli}.
\par
None of the above processes can be detected directly in a laboratory, but instead would produce
signals that travel through space to our detectors on or around Earth.
Knowledge of signal propagation is therefore a preliminary step for any phenomenological
analysis.
In this respect, the effective metric outside a compact spherical source derived from the bootstrapped   
Newtonian potential in four spacetime dimensions in Ref.~\cite{4Dmetric} is the simplest example
of what is necessary in order to confront with General Relativistic predictions.
Other cases are currently under investigation.
\par
Dimensional reduction offers novel phenomenology in many of these situations.
The small-mass ``catastrophe'' that plagues Hawking evaporation is replaced by a smooth,
thermodynamically-stable decay to a zero-mass remnant~\cite{cmn15}.
In fact, beyond gravitational physics, particle physics could also provide intriguing fingerprints.
The energy dependence of Drell-Yan scattering processes is modified as (energy)$^{-3}$ in $d=2$
spatial dimensions, as compared to (energy)$^{-2}$ in $d=3$, and the quadratic divergence of perturbative
Higgs mass corrections in $d=3$ spatial dimensions are tamed to linear and logarithmic ones in $d=2$ and $d=1$
respectively~\cite{stojkovic}.
In fact, pursuant to the current work underway with the effective bootstrapped Newtonian metric~\cite{4Dmetric},
a lower-dimensional version is currently in preparation~\cite{LowDmetric}.
\par
We stress that, like many current candidate quantum gravity theories, detecting these effects is problematic
at best.
Presently, there are no clear-cut predictions that single out the bootstrapped Newtonian approach,
but we stress that the results in this paper -- namely the cascading of dimensions and nested baby universes --
is reminiscent of existing work in the literature.
The novelty herein is that this behaviour has now been proposed in a non-linear Newtonian framework,
instead of the fully general relativistic one, but is nonetheless consistent with the demand that quantum
gravity naturally emerges as the number of spatial dimensions approaches one.
\section*{Acknowledgments}
R.C.~is partially supported by the INFN grant FLAG and his work has also been carried out in
the framework of activities of the National Group of Mathematical Physics (GNFM, INdAM)
and COST action {\it Cantata\/}. 
O.M.~is supported by the grant Laplas~VI of the Romanian National Authority for Scientific
Research. 
J.M.~thanks the Department of Physics and Astronomy at the University of Bologna for
its generous hospitality during the initial stages of this project.
\appendix
\section{Outer vacuum in $d \ge 3$}
\label{S:d>3}
\begin{figure}[t]
\centering
\includegraphics[width=8cm]{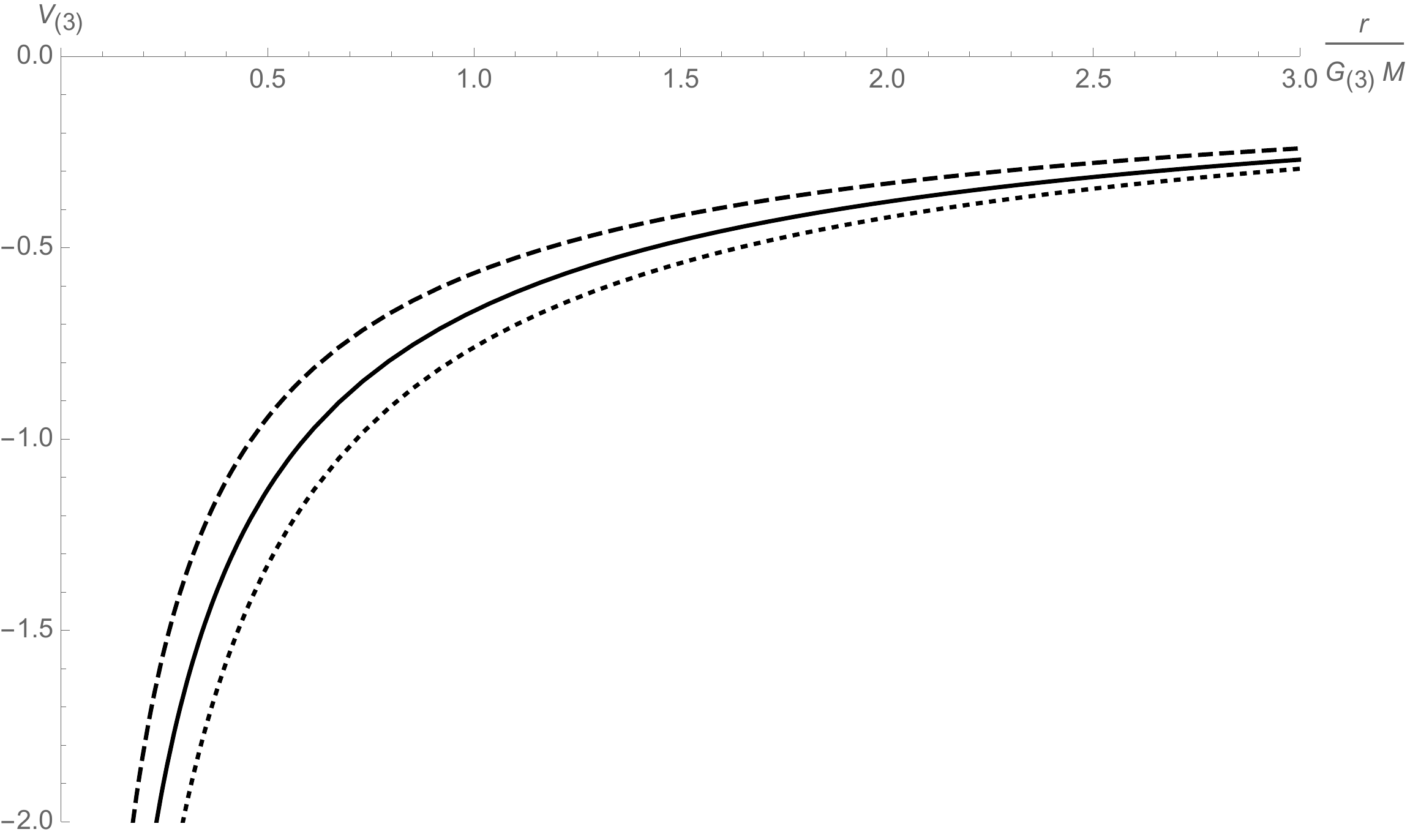}
$\ $
\includegraphics[width=8cm]{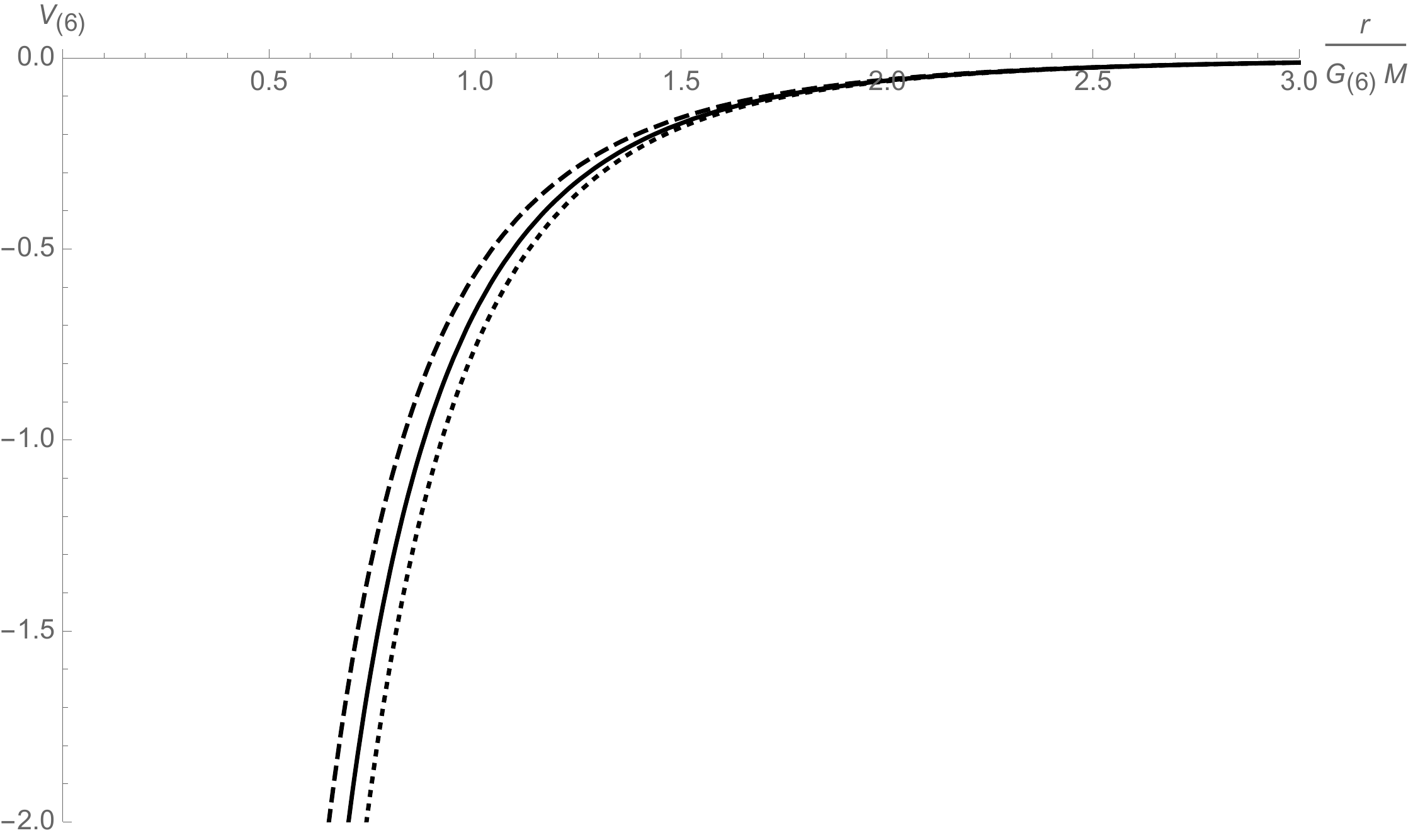}
\caption{Potential $V_{(3)}$ (left panel) and $V_{(6)}$ (right panel) for $q_V=1/2$ (dotted line), $q_V=1$ (solid line)
and $q_V=2$ (dashed line).}
\label{pV3-6}
\end{figure}

\setcounter{equation}{0}
In $d\ge 3$ spatial dimensions, the vacuum field equation~\eqref{EOMV0} is exactly solved by
\be
V_{(d\ge 3)}
=
\frac{\tilde V}{q_V}
=
\frac{1}{4\,q_V}
\left[
1-
\left(C_1+\frac{C_2}{r^{d-2}}\right)^{2/3}
\right]
\ ,
\ee
where the integration constants must be fixed in order to recover the Newtonian
behaviour at large distances,
\be
V_{\rm N}
=
-\frac{G_{(d)}\,M}{r^{d-2}}
\ ,
\label{Vn0}
\ee
that is $C_1=1$ and
\be
C_2
=
6\,q_V\,G_{(d)}\,M
\ .
\ee
These expressions finally yield
\be
V_{(d\ge 3)}
=
\frac{1}{4\,q_V}
\left[
1-
\left(1+\frac{6\,q_V\,G_{(d)}\,M}{r^{d-2}}\right)^{2/3}
\right]
\ .
\label{sol0}
\ee
Note that we can now take the limit $q_V\to 0$ and precisely recover the Newtonian potential~\eqref{Vn0},
as one would expect by first considering this limit in Eq.~\eqref{EOMV0}.
We also note that the large $r$ expansion of the solution~\eqref{sol0} reads
\be
V_{(d\ge 3)}
\simeq
-\frac{G_{(d)}\,M}{r^{d-2}}
+q_V\,\frac{G^2_{(d)}\,M^2}{r^{2\,(d-2)}}
\ ,
\ee
so that $q_V$ always affects the post-Newtonian order.
All solutions for $d\ge 3$ show essentially the same qualitative behaviour, with no restriction on the
range of possible values of $r$.
In Fig.~\ref{pV3-6} we plot the case $d=3$ and $d=6$ for different values of $q_V$ around one.
\end{document}